\documentstyle[12pt]{article}
\begin{document}
\begin{flushright}
hep-th/0311001\\
SNBNCBS-2003
\end{flushright}
\vskip 2.5cm
\begin{center}
{\bf \Large { Nilpotent Symmetries for QED in Superfield Formalism }}

\vskip 3cm

{\bf R.P.Malik}
\footnote{ E-mail address: malik@boson.bose.res.in  }\\
{\it S. N. Bose National Centre for Basic Sciences,} \\
{\it Block-JD, Sector-III, Salt Lake, Calcutta-700 098, India} \\

\vskip 2.5cm

\end{center}

\noindent
{\bf Abstract}:
In the framework of superfield approach, 
we derive the local, covariant, 
continuous and nilpotent (anti-)BRST and (anti-)co-BRST symmetry 
transformations on the $U(1)$ gauge field
($A_\mu$) and the (anti-)ghost fields ($(\bar C)C$) of the Lagrangian
density of the two ($1 + 1$)-dimensional QED by exploiting the 
(dual-)horizontality conditions defined on the 
four ($2 + 2)$-dimensional supermanifold. The long-standing problem of 
the derivation of the above symmetry transformations for the matter (Dirac) 
fields ($\bar \psi, \psi$) in the framework of superfield formulation is 
resolved by a new set of restrictions on 
the $(2 + 2)$-dimensional supermanifold. These new 
physically interesting restrictions on the supermanifold owe their 
origin to the invariance of conserved currents of the theory. 
The geometrical interpretation for all the above transformations is 
provided in the framework of superfield formalism.
\baselineskip=16pt

\vskip 1cm

\noindent
{\it PACS}: 11.15.-q; 12.20.-m; 11.30.Ph; 02.20.+b\\

\noindent
{\it Keywords}: Superfield formalism; (co-)BRST symmetries; 
                QED in two-dimensions\\

\newpage

\noindent
{\bf 1 Introduction}\\

\noindent
One of the most attractive and intuitive
geometrical approaches to gain an insight into
the physics and mathematics behind the
Becchi-Rouet-Stora-Tyutin (BRST) formalism is the superfield formulation
[1-6]. In this scheme, a $D$-dimensional
gauge theory (endowed with the first-class 
constraints in the language of Dirac [7,8]) is considered on a 
$(D + 2)$-dimensional supermanifold parameterized by $D$-number of spacetime
(even) co-ordinates $x^\mu$ ($\mu = 0, 1, 2....D-1$) and a couple of
(odd) Grassmannian variables $\theta$ and $\bar\theta$ (with $\theta^2
= \bar\theta^2 = 0, \theta\bar\theta + \bar\theta \theta = 0$). In general,
the $(p + 1)$-form super curvature $\tilde F$ constructed from the super
exterior derivative $\tilde d$ (with $\tilde d^2 = 0$) and the super
$p$-form connection $\tilde A$ of a $p$-form ($p = 1, 2, 3....$) 
gauge theory through the Maurer-Cartan equation (i.e. $\tilde d \tilde A
+ \tilde A \wedge \tilde A = \tilde F $)
is restricted to be flat along the Grassmannian directions
of the $(D + 2)$-dimensional supermanifold due
to the so-called horizontality condition
\footnote{ Nakanishi and Ojima call it the ``soul-flatness'' condition
which amounts to setting the Grassmannian components of a 
$(p + 1)$-form super curvature tensor (for a $p$-form
gauge theory) equal to zero [9].}. Mathematically,
this condition implies $\tilde F = F$ where
$F = d A + A \wedge A$ is the $(p + 1)$-form curvature defined on the
ordinary $D$-dimensional spacetime manifold. The
horizontality condition, where
only one of the three de Rham cohomological operators
\footnote{ On an ordinary manifold 
without a boundary, the three operators $(d, \delta,\Delta)$
form a set of de Rham cohomological operators where $(\delta)d$ are the
(co-)exterior derivatives with $d = dx^\mu \partial_\mu, \delta = \pm * d *$
and $d^2 = \delta^2 = 0$. Here $*$ is the Hodge duality operation on the
manifold. The Laplacian operator $\Delta = (d + \delta)^2 = \{ d, \delta \}$
turns out to be the Casimir operator for the full set of algebra:
$\delta^2 = 0, d^2 = 0, \Delta = \{d , \delta \}, [\Delta, d ] = 0,
[\Delta, \delta] = 0$ obeyed by these cohomological operators belonging
to the geometrical aspects of the subject of differential geometry
(see, e.g., [10-14] for details).} is exploited, leads to the 
derivation of the nilpotent (anti-)BRST symmetry transformations on
the gauge- and (anti-)ghost fields of the (anti-)BRST invariant Lagrangian
density of a given $D$-dimensional $p$-form gauge theory.

In a recent set of papers [15-17], 
all the three (super) de Rham cohomological operators have been exploited,
in the generalized versions of the horizontality condition,
to derive the (anti-)BRST, (anti-)co-BRST and a bosonic symmetry
(which is equal to the anticommutator(s) of the (anti-)BRST and (anti-)co-BRST 
symmetries) transformations
for the free one-form Abelian gauge theory in two-dimensions (2D) of spacetime.
For the derivation of the above nilpotent symmetries, the super (co-)exterior
derivatives $(\tilde \delta)\tilde d$ have been exploited in the
(dual-)horizontality conditions on the four $(2 + 2)$-dimensional
supermanifold. The Lagrangian formulation of the above symmetries
has also been carried out in a set of papers [18-20] where it has been shown 
that this theory presents (i) an example of a tractable field theoretical
model for the Hodge theory, and (ii) an example of a new class of
topological field theory where the Lagrangian density turns out to be like 
Witten type topological field theory but the symmetries of the theory
are that of Schwarz type.
Similar symmetries for the self-interacting 2D non-Abelian gauge theory
have also been obtained in the framework of 2D Lagrangian formalism [21]
as well as in the four $(2 + 2)$-dimensional superfield formulation [22].
Furthermore, the above type of symmetries have been shown
to exist for the 4D
2-form free Abelian gauge theory in the Lagrangian formalism [23,24].

One of the most difficult and long-standing problems in the realm of
superfield approach to BRST formalism has been to derive the
(anti-)BRST symmetry transformations on the matter (e.g. Dirac,
complex scalar etc.) fields for a given interacting $p$-form gauge theory.
The purpose of the present paper is to demonstrate that an additional set
of restrictions, besides the (dual-)horizontality conditions w.r.t 
super (co-)exterior derivatives $(\tilde \delta) \tilde d$,
 are required on the $(D + 2)$-dimensional supermanifold
for the derivation of the (anti-)BRST and (anti-)co-BRST transformations
on the matter fields. For this purpose, as a prototype field theoretical
model, we choose the two-dimensional
interacting $U(1)$ gauge theory (i.e. QED
\footnote{ A dynamically 
closed and locally gauge invariant
system of the photon and Dirac fields.}) 
and show that the
(anti-)BRST and (anti-)co-BRST symmetry transformations on the matter fields,
derived in our earlier works [25,26] in the framework of Lagrangian formalism,
can be obtained by exploiting the invariance of the conserved (super) currents
constructed by the (super) Dirac fields of the theory
on a (super)manifold. In a more precise and sophisticated
language, the {\it equality} of the supercurrents 
$\tilde J_\mu (x,\theta,\bar\theta)$ and
$\tilde J^{(5)}_\mu (x,\theta,\bar\theta)$ constructed by the superfields 
(cf. eqns. (4.2) and (4.9) below) on
the four $(2 + 2)$-dimensional supermanifold {\it with} the conserved currents 
$J_\mu (x) = (\bar\psi \gamma_\mu \psi)(x)$ and 
$J^{(5)}_\mu ((x) = (\bar\psi \gamma_\mu\gamma_5 \psi)(x)$ constructed
by the ordinary Dirac fields on the 2D ordinary manifold 
leads to the derivation  of the (anti-)BRST 
and (anti-)co-BRST symmetry transformations on the 
Dirac fields, respectively. The above equality emerges automatically 
and is not imposed by hand. We also provide, in the present paper, 
the geometrical interpretations for the nilpotent symmetries and
the corresponding nilpotent generators.

The outline of our present paper is as follows. In Section 2, we 
recapitulate the salient features of our earlier works [25,26]
on the existence of the off-shell nilpotent (anti-)BRST- and (anti-)co-BRST
symmetries in the Lagrangian formulation for the {\it interacting} $U(1)$
gauge theory in two-dimensions of spacetime. Section 3 is devoted to
the derivation of the above symmetry transformations on the gauge field
$A_\mu$ and the (anti-)ghost fields $(\bar C)C$ by exploiting the
(dual-)horizontality conditions on the four $(2 + 2)$-dimensional
 supermanifold [17,22]. This exercise is carried
out for the sake of this paper to be self-contained. 
The central of our paper is Section 4 where we derive
the above symmetry transformations for the matter (Dirac) fields by
invoking the {\it invariance} of the conserved currents as the physical 
restriction on the supermanifold. Finally, we make some concluding remarks 
and pinpoint a few future directions in Section 5 for further investigations.\\

\noindent
{\bf 2 Preliminary: (anti-)BRST- and (anti-)co-BRST symmetries}\\

\noindent
To recapitulate the bare essentials of our earlier works [25,26] 
on QED in two-dimensions,
let us begin with the (anti-)BRST invariant Lagrangian density ${\cal L}_{b}$
for the {\it interacting} two ($1 + 1)$-dimensional (2D) $U(1)$ gauge theory
in the Feynman gauge [27-29]
$$
\begin{array}{lcl}
{\cal L}_{b} &=& - \frac{1}{4}\; F^{\mu\nu} F_{\mu\nu} 
+ \bar \psi \;(i \gamma^\mu D_\mu - m)\; \psi + B \;(\partial \cdot A)
+ \frac{1}{2}\; B^2
- i \;\partial_{\mu} \bar C \partial^\mu C \nonumber\\
&\equiv& \frac{1}{2}\; E^2
+ \bar \psi\; (i \gamma^\mu D_\mu - m) \;\psi + B \;(\partial \cdot A)
+ \frac{1}{2}\; B^2
- i \;\partial_{\mu} \bar C \partial^\mu C, 
\end{array} \eqno(2.1)
$$
where $F_{\mu\nu} = \partial_\mu A_\nu - \partial_\nu A_\mu$ is the field 
strength tensor for the $U(1)$ gauge theory that is derived from the 2-form
$d A = \frac{1}{2} (dx^\mu \wedge dx^\nu) F_{\mu\nu}$
\footnote{We adopt here the conventions and notations such that the 2D flat
Minkowski metric is: $\eta_{\mu\nu} =$ diag $(+1, -1)$ and $\Box = 
\eta^{\mu\nu} \partial_{\mu} \partial_{\nu} = (\partial_{0})^2 - 
(\partial_{1})^2, \varepsilon_{\mu\nu} = - \varepsilon^{\mu\nu}, F_{01} 
= E = \partial_{0} A_{1} - \partial_{1} A_{0} = - \varepsilon^{\mu\nu}
\partial_\mu A_\nu = F^{10}, \varepsilon_{01} =
\varepsilon^{10} = + 1, \;D_{\mu} \psi = \partial_{\mu} \psi + i e 
A_{\mu} \psi$. The Dirac $\gamma$ matrices in two-dimensions are chosen to be:
$\gamma^0 = \sigma_2, \gamma^1 = i \sigma_1, \gamma_5 = \gamma^0 \gamma^1
= \sigma_3, \{\gamma^\mu, \gamma^\nu \} = 2 \eta^{\mu\nu}, \gamma_\mu \gamma_5
= \varepsilon_{\mu\nu} \gamma^\nu$. Here $\sigma$'s are the usual
$2 \times 2$ Pauli matrices and the Greek indices: $\mu, \nu, \rho...
= 0, 1$ correspond to the spacetime directions on the manifold.}. 
As is evident, the latter
is  constructed by the application of the exterior derivative 
$d = dx^\mu \partial_\mu$ (with $d^2 = 0)$ on the 1-form $A = dx^\mu A_\mu$
(which defines the vector potential $A_\mu$). It will be noted that in 2D,
$F_{\mu\nu}$ has only the electric component (i.e. $F_{01} = E$) and there
is no magnetic component associated with it. The
gauge-fixing term $(\partial \cdot A)$ is derived through the operation
of the co-exterior derivative $\delta$ 
(with $\delta = - * d *, \delta^2 = 0$) on the
one-form $A$ (i.e. $\delta A = - * d * A = (\partial \cdot A)$)
where $*$ is the Hodge duality operation. The fermionic
Dirac fields $(\psi, \bar \psi)$, with the mass $m$ and charge $e$, couple 
to the $U(1)$ gauge field $A_\mu$ (i.e. $ - e \bar \psi \gamma^\mu A_\mu \psi$) through the conserved current 
$J_\mu = \bar \psi \gamma_\mu \psi$. The
anticommuting ($ C \bar C + \bar C C = 0, C^2 = \bar C^2 = 0,
C \psi + \psi C = 0$ etc.) (anti-)ghost fields $(\bar C)C$ are required to
maintain the unitarity and ``quantum'' gauge (i.e. BRST) invariance together
at any arbitrary order of perturbation theory
\footnote{ The full strength of the (anti-)ghost fields turns up in the
discussion of the unitarity and gauge invariance
for the perturbative computations in the realm of non-Abelian gauge theory
where the loop diagrams of the gauge (gluon) fields play 
a very important role
(see, e.g., [30] for details).}. The kinetic energy term 
$(\;\frac{1}{2} E^2\;)$
of (2.1) can be linearized by invoking an auxiliary field ${\cal B}$
$$
\begin{array}{lcl}
{\cal L}_{B} =  {\cal B} \; E - \frac{1}{2} {\cal B}^2
+ \bar \psi\; (i \gamma^\mu D_\mu - m)\; \psi + B\; (\partial \cdot A)
+ \frac{1}{2}\; B^2
- i\; \partial_{\mu} \bar C \partial^\mu C,
\end{array} \eqno(2.2)
$$
which is the analogue of the Nakanishi-Lautrup auxiliary field $B$ that is
required to linearize the gauge-fixing term $-\frac{1}{2} (\partial\cdot A)^2$
in (2.1).
The above Lagrangian density
(2.2) respects the following off-shell nilpotent
$(s_{(a)b}^2 = 0,  s_{(a)d}^2 = 0)$ (anti-)BRST ($s_{(a)b}$)
\footnote{We adopt here the notations and conventions followed in [29]. 
In fact, in its full glory, a nilpotent ($\delta_{B}^2 = 0$)
BRST transformation $\delta_{B}$ is equivalent to the product of an 
anticommuting ($\eta C = - C \eta, \eta \bar C = - \bar C\eta,
\eta \psi = - \psi \eta, \eta \bar \psi = - \bar \psi \eta$ etc.)
spacetime independent parameter $\eta$ and $s_{b}$ 
(i.e. $\delta_{B} = \eta \; s_{b}$) where $s_{b}^2 = 0$.} 
- and (anti-)dual(co)-BRST ($s_{(a)d}$) symmetry transformations 
(with $s_b s_{ab} + s_{ab} s_b = 0, s_d s_{ad} + s_{ad} s_d = 0$) [25,26]
$$
\begin{array}{lcl}
s_{b} A_{\mu} &=& \partial_{\mu} C, \qquad 
s_{b} C = 0, \qquad 
s_{b} \bar C = i B,  \qquad s_b \psi = - i e C \psi, \nonumber\\
s_b \bar \psi &=& - i e \bar \psi C,
\qquad s_{b} {\cal B} = 0, \quad  s_{b} B = 0, \quad
\;s_{b} E = 0, \quad s_b (\partial \cdot A) = \Box C, \nonumber\\
s_{ab} A_{\mu} &=& \partial_{\mu} \bar C, \qquad 
s_{ab} \bar C = 0, \qquad 
s_{ab} C = - i B,  \qquad s_{ab} \psi = - i e \bar C \psi, \nonumber\\
s_{ab} \bar \psi &=& - i e \bar \psi \bar C,
\qquad s_{ab} {\cal B} = 0, \quad  s_{ab} B = 0, \quad
\;s_{ab} E = 0, \quad s_{ab} (\partial \cdot A) = \Box \bar C, 
\end{array}\eqno(2.3)
$$
$$
\begin{array}{lcl}
s_{d} A_{\mu} &=& - \varepsilon_{\mu\nu} \partial^\nu \bar C, \quad
\quad s_{d} B = 0, \quad s_{d} (\partial \cdot A) = 0, \quad
s_{d} \bar C = 0, \quad s_{d} C = - i {\cal B}, \nonumber\\
s_{d} {\cal B} &=& 0, \qquad s_d \psi = - i e \bar C \gamma_5 \psi,
\qquad s_d \bar \psi = + i e \bar \psi \bar C \gamma_5,
\qquad s_{d} E = \Box \bar C, \nonumber\\
s_{ad} A_{\mu} &=& - \varepsilon_{\mu\nu} \partial^\nu  C, \quad
\quad s_{ad} B = 0, \quad s_{ad} (\partial \cdot A) = 0, \quad
s_{ad}  C = 0, \quad s_{ad} \bar C = + i {\cal B}, \nonumber\\
s_{ad} {\cal B} &=& 0, \qquad s_{ad} \psi = - i e C \gamma_5 \psi,
\qquad s_{ad} \bar \psi = + i e \bar \psi  C \gamma_5
\qquad s_{ad} E = \Box  C.
\end{array}\eqno(2.4)
$$
The noteworthy points, at this stage, are (i) under the (anti-)BRST
and (anti-)co-BRST transformations, it is the kinetic energy term 
(more precisely $E$ itself) and the gauge-fixing term
(more accurately $(\partial \cdot A)$ itself) that remain invariant,
respectively. (ii) The electric field $E$ and $(\partial \cdot A)$
owe their origin to the operation of
cohomological operators $d$ and $\delta$ on the one-form $A = dx^\mu A_\mu$,
respectively. (iii) For the (anti-)co-BRST transformations to be the
symmetry transformations for (2.2), there exists the restriction that
$m = 0$ for the Dirac fields. There is no such restriction for the
validity of the (anti-)BRST symmetry transformations. (iv) The 
anticommutator $(s_w = \{ s_{b} s_{d} \} = \{ s_{ab}, s_{ad} \})$
of the above nilpotent symmetries is a bosonic symmetry 
transformation $s_w$ (with
$s_w^2 \neq 0$) for the Lagrangian density (2.2) [26]. 
(v) The operator algebra among the above transformations is exactly
identical to the algebra obeyed by the de Rham cohomological operators.
(vi) The symmetry transformations in (2.3) and (2.4) are generated  by
the local, conserved and nilpotent charges $Q_{(a)b}$ and $Q_{(a)d}$. This
statement can be succinctly expressed in the mathematical form as
$$
\begin{array}{lcl}
s_{r}\; \Sigma (x) = - i\; 
\bigl [\; \Sigma (x),  Q_r\; \bigr ]_{\pm}, \qquad\;\;\;
r = b, ab, d, ad,
\end{array} \eqno(2.5)
$$
where the local generic field
$\Sigma = A_\mu, C, \bar C, \psi, \bar \psi, B, 
{\cal B}$ and the $(+)-$ signs, 
as the subscripts on the (anti-)commutator $[\;, \;]_{\pm}$, 
stand for $\Sigma$ being (fermionic)bosonic in nature.\\

\noindent
{\bf 3 Nilpotent symmetries for the gauge- and (anti-)ghost fields}\\

\noindent
We begin here with a four ($2 + 2$)-dimensional supermanifold
parametrized by the superspace coordinates $Z^M = (x^\mu, \theta, \bar \theta)$
where $x^\mu\; (\mu = 0, 1)$ are a couple of
even (bosonic) spacetime coordinates
and $\theta$ and $\bar \theta$ are the two odd (Grassmannian) coordinates
(with $\theta^2 = \bar \theta^2 = 0, 
\theta \bar \theta + \bar \theta \theta = 0)$. On this supermanifold, one can
define a supervector superfield $\tilde A_M$ (i.e.
$\tilde A_M = ( B_{\mu} (x, \theta, \bar \theta), 
\;\Phi (x, \theta, \bar \theta),
\;\bar \Phi (x, \theta, \bar \theta))$
with $B_\mu, \Phi, \bar \Phi$ as the component
multiplet superfields [4]. The superfields $B_{\mu}, \Phi, \bar \Phi $ 
can be expanded in terms
of the basic fields ($A_\mu, C, \bar C$) and  auxiliary fields
($B, {\cal B}$) of  (2.2) and some extra secondary fields as follows
$$
\begin{array}{lcl}
B_{\mu} (x, \theta, \bar \theta) &=& A_{\mu} (x) 
+ \theta\; \bar R_{\mu} (x) + \bar \theta\; R_{\mu} (x) 
+ i \;\theta \;\bar \theta S_{\mu} (x), \nonumber\\
\Phi (x, \theta, \bar \theta) &=& C (x) 
+ i\; \theta \bar B (x)
- i \;\bar \theta\; {\cal B} (x) 
+ i\; \theta\; \bar \theta \;s (x), \nonumber\\
\bar \Phi (x, \theta, \bar \theta) &=& \bar C (x) 
- i \;\theta\;\bar {\cal B} (x) + i\; \bar \theta \;B (x) 
+ i \;\theta \;\bar \theta \;\bar s (x).
\end{array} \eqno(3.1)
$$
It is straightforward to note that the local 
fields $ R_{\mu} (x), \bar R_{\mu} (x),
C (x), \bar C (x), s (x), \bar s (x)$ are fermionic (anti-commuting) 
in nature and the bosonic (commuting) local fields in (3.1)
are: $A_{\mu} (x), S_{\mu} (x), {\cal B} (x), \bar {\cal B} (x),
B (x), \bar B (x)$. It is unequivocally clear
that, in the above expansion, the bosonic-
 and fermionic degrees of freedom match. This requirement is essential
for the validity and sanctity of any arbitrary supersymmetric theory in the 
superfield formulation. In fact, all the secondary fields will be expressed 
in terms of basic fields due to the restrictions emerging from the application 
of horizontality condition (i.e. $\tilde F = F$), namely;
$$
\begin{array}{lcl} 
\tilde F =  \frac{1}{2}\; (d Z^M \wedge d Z^N)\;
\tilde F_{MN} = \tilde d \tilde A  \equiv
d A = \frac{1}{2} (dx^\mu \wedge dx^\nu)\; F_{\mu\nu} = F,
\end{array} \eqno(3.2)
$$
where the super exterior derivative $\tilde d$ and 
the connection super one-form $\tilde A$ are defined as
$$
\begin{array}{lcl}
\tilde d &=& \;d Z^M \;\partial_{M} = d x^\mu\; \partial_\mu\;
+ \;d \theta \;\partial_{\theta}\; + \;d \bar \theta \;\partial_{\bar \theta},
\nonumber\\
\tilde A &=& d Z^M\; \tilde A_{M} = d x^\mu \;B_{\mu} (x , \theta, \bar \theta)
+ d \theta\; \bar \Phi (x, \theta, \bar \theta) + d \bar \theta\;
\Phi ( x, \theta, \bar \theta).
\end{array}\eqno(3.3)
$$
In physical language, this requirement implies that the physical field
$E$, derived from the curvature term $F_{\mu\nu}$, does not get any
contribution from the Grassmannian variables. In other words, the
physical electric field $E$ for 2D QED remains intact in the
superfield formulation. Mathematically, the condition (3.2) implies
the ``flatness'' of all the components of the
super curvature (2-form) tensor $\tilde F_{MN}$ that are directed along the 
 $\theta$ and/or $\bar \theta$ directions of the supermanifold. To this
end in mind, first we expand $\tilde d \tilde A$ as
$$
\begin{array}{lcl}
\tilde d \tilde A &=& (d x^\mu \wedge d x^\nu)\;
(\partial_{\mu} B_\nu) - (d \theta \wedge d \theta)\; (\partial_{\theta}
\bar \Phi) + (d x^\mu \wedge d \bar \theta)
(\partial_{\mu} \Phi - \partial_{\bar \theta} B_{\mu}) \nonumber\\
&-& (d \theta \wedge d \bar \theta) (\partial_{\theta} \Phi 
+ \partial_{\bar \theta} \bar \Phi) 
+ (d x^\mu \wedge d \theta) (\partial_{\mu} \bar \Phi - \partial_{\theta}
B_{\mu}) - (d \bar \theta \wedge d \bar \theta)
(\partial_{\bar \theta} \Phi). 
\end{array}\eqno(3.4)
$$
Ultimately, the application of soul-flatness (horizontality) condition
($\tilde d \tilde A = d A$) yields [17]
$$
\begin{array}{lcl}
R_{\mu} \;(x) &=& \partial_{\mu}\; C(x), \qquad 
\bar R_{\mu}\; (x) = \partial_{\mu}\;
\bar C (x), \qquad \;s\; (x) = \bar s\; (x) = 0,
\nonumber\\
S_{\mu}\; (x) &=& \partial_{\mu} B\; (x) 
\qquad
B\; (x) + \bar B \;(x) = 0, \qquad 
{\cal B}\; (x)  = \bar {\cal B} (x) = 0.
\end{array} \eqno(3.5)
$$
The insertion of all the above values in the expansion (3.1) leads to
the derivation of the (anti-)BRST symmetries for the 
gauge- and (anti-)ghost fields of the Abelian gauge theory.
In addition, this exercise provides  the physical interpretation for the
(anti-)BRST charges $Q_{(a)b}$ 
as the generators (cf. eqn. (2.5)) of translations 
(i.e. $ \mbox{Lim}_{\bar\theta \rightarrow 0} (\partial/\partial \theta),
 \mbox{Lim}_{\theta \rightarrow 0} (\partial/\partial \bar\theta)$)
along the Grassmannian
directions of the supermanifold. Both these observations can be succinctly 
expressed, in a combined way, by re-writing the super expansion (3.1) as
$$
\begin{array}{lcl}
B_{\mu}\; (x, \theta, \bar \theta) &=& A_{\mu} (x) 
+ \;\theta\; (s_{ab} A_{\mu} (x)) 
+ \;\bar \theta\; (s_{b} A_{\mu} (x)) 
+ \;\theta \;\bar \theta \;(s_{b} s_{ab} A_{\mu} (x)), \nonumber\\
\Phi\; (x, \theta, \bar \theta) &=& C (x) \;+ \; \theta\; (s_{ab} C (x))
\;+ \;\bar \theta\; (s_{b} C (x)) 
\;+ \;\theta \;\bar \theta \;(s_{b}\; s_{ab} C (x)), 
 \nonumber\\
\bar \Phi\; (x, \theta, \bar \theta) &=& \bar C (x) 
\;+ \;\theta\;(s_{ab} \bar C (x)) \;+\bar \theta\; (s_{b} \bar C (x))
\;+\;\theta\;\bar \theta \;(s_{b} \;s_{ab} \bar C (x)).
\end{array} \eqno(3.6)
$$

To obtain the (anti-)co-BRST transformations on the gauge- and (anti-)ghost
fields, we exploit the dual-horizontality condition
$\tilde \delta \tilde A = \delta A$ on the $(2 + 2)$-dimensional
supermanifold where $\tilde \delta = - \star\; \tilde d \; \star$ is the super co-exterior derivative on the four $(2 + 2)$-dimensional supermanifold
and $\delta = - * d *$ is the co-exterior derivative on the ordinary 2D
manifold. The Hodge duality operations on the supermanifold and ordinary
manifold are denoted by $\star$ and $*$, respectively. The $\star$ operations
on the super differentials $(dZ^M)$ and their wedge products
$(dZ^M \wedge dZ^N)$, etc., defined on the $(2 + 2)$-dimensional
supermanifold, are [22,31]
$$
\begin{array}{lcl}
&&\star\; (dx^\mu) = \varepsilon^{\mu\nu}\; 
(dx_\nu \wedge d \theta \wedge d\bar\theta), \;\;\;\;\;\qquad\;\;\;
\star\; (d\theta) = \frac{1}{2!}\;\varepsilon^{\mu\nu}\; 
(dx_\mu \wedge dx_\nu \wedge d\bar\theta), \nonumber\\
&&\star\; (d\bar\theta) = \frac{1}{2!}\;\varepsilon^{\mu\nu}\; 
(dx_\mu \wedge dx_\nu \wedge d\theta), \;\;\qquad\;\;
\star\; (dx^\mu \wedge dx^\nu) = \varepsilon^{\mu\nu}\; 
(d \theta \wedge d\bar\theta), \nonumber\\
&&\star\; (dx^\mu \wedge d\theta) = \varepsilon^{\mu\nu}\; 
(d x_\nu \wedge d\bar\theta), \;\;\;\qquad\;\;\;\;\;
\star\; (dx^\mu \wedge d\bar\theta) = \varepsilon^{\mu\nu}\; 
(d x_\nu \wedge d\theta), \nonumber\\
&&\star\; (d\theta \wedge d\theta) = \frac{1}{2!}
s^{\theta\theta}\;\varepsilon^{\mu\nu}\; 
(dx_\mu \wedge d x_\nu), \qquad
\star\; (d\theta \wedge d\bar\theta) = \frac{1}{2!}\; 
\varepsilon^{\mu\nu}\; 
(d x_\mu \wedge d x_\nu), \nonumber\\
&&\star\; (d \bar\theta \wedge d\bar\theta) = \frac{1}{2!}
s^{\bar\theta\bar\theta}\;\varepsilon^{\mu\nu}\; 
(dx_\mu \wedge dx_\nu), \;\;\quad\;\; 
\star\;(dx_\mu \wedge d \theta \wedge d\bar\theta) = \varepsilon_{\mu\nu}
(d x^\nu), \nonumber\\
&&\star\;(dx_\mu \wedge dx_\nu
\wedge d \theta \wedge d\bar\theta) = \varepsilon_{\mu\nu}, 
\;\;\;\qquad\;\;\;\;\;
\star\; (dx_\mu \wedge dx_\nu \wedge d \theta) = 
\varepsilon_{\mu\nu} (d \bar \theta), \nonumber\\
&&\star\; (dx_\mu \wedge dx_\nu \wedge d \bar\theta) = 
\varepsilon_{\mu\nu} (d \theta), \;\;\;\;\;\qquad\;\;\;\;
\star \; (dx_\mu \wedge dx_\nu \wedge d\theta \wedge d\theta)
= \varepsilon_{\mu\nu} s^{\theta\theta},
\end{array} \eqno(3.7)
$$
where $s$'s are the symmetric  {\it constant} quantities on the Grassmannian
submanifold of the four ($2 + 2$)-dimensional supermanifold. They are
introduced to take care of the fact that two successive $\star$
operation on any differential should yield the same differential
(see, [31] for detail discussions). 
With the above inputs,
it can be checked that the superscalar superfield $\tilde \delta \tilde A
= - \star \tilde d  \star \tilde A$, 
 turns out to be
$$
\begin{array}{lcl}
\tilde \delta \tilde A = (\partial \cdot B 
+ \partial_{\theta} \bar \Phi + \partial_{\bar\theta} \Phi)
+ s^{\theta\theta}
\;(\partial_{\theta} \Phi) + s^{\bar\theta\bar\theta}\;
(\partial_{\bar\theta} \bar \Phi). 
\end{array} \eqno(3.8)
$$
Ultimately, the dual-horizontality
restriction $\tilde \delta \tilde A = \delta A$ produces the following
restrictions on the component superfields (see, e.g., [31] for details)
$$
\begin{array}{lcl}
\partial_\theta \Phi = 0, \qquad \partial_{\bar\theta} \bar\Phi = 0,
\qquad (\partial \cdot B + \partial_{\bar\theta} \Phi + \partial_{\theta}
\bar \Phi)
= (\partial \cdot A),
\end{array} \eqno(3.9)
$$
where, as is evident, the r.h.s. of the last entry in the above equation
is due to $\delta A = (\partial \cdot A)$. Exploiting the super 
expansions of (3.1), we obtain
$$
\begin{array}{lcl}
&&(\partial \cdot R) (x) = (\partial \cdot \bar R) (x) 
= (\partial \cdot S) (x) = 0,
\qquad s\; (x) = \bar s \; (x) = 0, \nonumber\\
&& B\; (x) = 0, \quad \bar B \;(x) = 0, \;\;\;\qquad\;\;\;
{\cal B}\; (x) + \bar {\cal B}\; (x) = 0.
\end{array} \eqno(3.10)
$$
It is clear from the above 
that we cannot get a {\it unique} solution for $R_\mu, \bar R_\mu$
and $S_\mu$ in terms of the basic fields of the Lagrangian density (2.2). This
is why there are non-local and non-covariant
solutions for these in the case of QED in 4D (see, e.g., [31]).
It is interesting, however, to point out that for 
2D QED, we have the local and covariant solutions as
$$
\begin{array}{lcl}
R_\mu = - \varepsilon_{\mu\nu} \partial^\nu \bar C, 
\qquad \bar R_\mu = - \varepsilon_{\mu\nu} \partial^\nu C, \qquad
S_\mu = + \varepsilon_{\mu\nu} \partial^\nu {\cal B}.
\end{array} \eqno(3.11)
$$
With the above insertions, it can be easily checked that the expansion (3.1)
becomes
$$
\begin{array}{lcl}
B_{\mu}\; (x, \theta, \bar \theta) &=& A_{\mu} (x) 
+ \;\theta\; (s_{ad} A_{\mu} (x)) 
+ \;\bar \theta\; (s_{d} A_{\mu} (x)) 
+ \;\theta \;\bar \theta \;(s_{d} s_{ad} A_{\mu} (x)), \nonumber\\
\Phi\; (x, \theta, \bar \theta) &=& C (x) \;+ \; \theta\; (s_{ad} C (x))
\;+ \;\bar \theta\; (s_{d} C (x)) 
\;+ \;\theta \;\bar \theta \;(s_{d}\; s_{ad} C (x)), 
 \nonumber\\
\bar \Phi\; (x, \theta, \bar \theta) &=& \bar C (x) 
\;+ \;\theta\;(s_{ad} \bar C (x)) \;+\bar \theta\; (s_{d} \bar C (x))
\;+\;\theta\;\bar \theta \;(s_{d} \;s_{ad} \bar C (x)).
\end{array} \eqno(3.12)
$$
Thus, the geometrical interpretation for the generators $Q_{(a)d}$ of
the (anti-)co-BRST symmetries is identical to that of the (anti-)BRST
charges $Q_{(a)b}$. However, there is a clear-cut distinction between
$Q_{(a)d}$ and $Q_{(a)b}$ when the transformations on the (anti-)ghost
fields are considered. For instance, the BRST charge $Q_b$ generates
a symmetry transformation such that the  superfield 
$\bar\Phi (x,\theta,\bar\theta)$ becomes {\it anti-chiral} and the 
superfield  $\Phi (x,\theta,\bar\theta)$ becomes an 
ordinary local field $C(x)$. 
In contrast, the co-BRST charge $Q_d$ generates
a symmetry transformation under which just the 
{\it opposite} of the above happens. 
Similarly, the distinction between
$Q_{ab}$ and $Q_{ad}$ can be argued where one of the above superfields
becomes {\it chiral}.\\

\noindent
{\bf 4 Nilpotent symmetries for the Dirac fields}\\

\noindent
In contrast to the (dual-)horizontality conditions that rely on the
(super-)co-exterior derivatives $(\tilde \delta)\delta$, the
(super-)exterior derivative $(\tilde d)d$ and the (super-)one-form
$(\tilde A)A$ for the derivation of the (anti-)BRST and (anti-)co-BRST 
symmetry transformations  on the gauge field $A_\mu$
and the (anti-)ghost fields $(\bar C)C$, the 
corresponding nilpotent symmetries
for the matter (Dirac) fields $(\psi, \bar\psi)$ are obtained due to the
invariance of the conserved currents of the theory. To corroborate this
assertion, first of all, we start off with the super expansion of the
superfields $(\Psi, \bar\Psi)(x, \theta,\bar\theta)$),
corresponding to the ordinary Dirac fields $(\psi, \bar\psi)(x)$,
as
$$
\begin{array}{lcl} 
 \Psi (x, \theta, \bar\theta) &=& \psi (x)
+ i \;\theta\; \bar b_1 (x) + i \;\bar \theta \; b_2 (x) 
+ i \;\theta \;\bar \theta \;f (x),
\nonumber\\
\bar \Psi (x, \theta, \bar\theta) &=& \bar \psi (x)
+ i\; \theta \;\bar b_2 (x) + i \;\bar \theta \; b_1 (x) 
+ i\; \theta \;\bar \theta \;\bar f (x).
\end{array} \eqno(4.1)
$$
It is clear and evident that, in the limit 
$(\theta, \bar\theta) \rightarrow 0$,
we get back the Dirac fields $(\psi, \bar\psi)$ of the 
Lagrangian density (2.1). Furthermore, the number of
bosonic fields ($b_1, \bar b_1, b_2, \bar b_2)$ match with the fermionic
fields $(\psi, \bar \psi, f, \bar f)$ so that the above expansion is consistent
with the basic tenets of supersymmetry. Now one can construct the
supercurrent $\tilde J_\mu (x, \theta, \bar\theta)$ from the above
superfields with the following general super expansion
$$
\begin{array}{lcl} 
\tilde J_\mu (x, \theta, \bar\theta) = \bar \Psi (x,\theta,\bar\theta)
\;\gamma_\mu \;\Psi (x, \theta, \bar\theta)
= J_\mu (x) + \theta \; \bar K_\mu (x)
+ \bar \theta\; K_\mu (x) + i \; \theta\; \bar\theta\; L_\mu (x), 
\end{array} \eqno(4.2)
$$
where the above components (i.e. $\bar K_\mu, K_\mu, L_\mu, J_\mu$),
along the  Grassmannian directions
$\theta$ and  $\bar\theta$ as well as the bosonic directions
$\theta\bar\theta$ and identity $\hat {\bf 1}$ of the
supermanifold, can be expressed in terms of the components of the
basic super expansions (4.1), as
$$
\begin{array}{lcl} 
&& \bar K_\mu (x) = i \bigl ( \bar b_2 \gamma_\mu \psi -
\bar \psi \gamma_\mu \bar b_1 \bigr ),
\qquad  K_\mu (x) = i \bigl ( b_1 \gamma_\mu \psi -
\bar \psi \gamma_\mu  b_2 \bigr ), \nonumber\\
&& L_\mu (x) = \bar f \gamma_\mu \psi + \bar \psi \gamma_\mu f
+ i (\bar b_2 \gamma_\mu b_2 - b_1 \gamma_\mu \bar b_1), 
\qquad J_\mu (x) = \bar \psi  \gamma_\mu \psi.
 \end{array} \eqno(4.3)
$$
To be consistent with our earlier observation that the (co-)BRST
transformations $(s_{(d)b})$ are equivalent to the translations
(i.e. $\mbox{Lim}_{\theta \rightarrow 0} (\partial/\partial \bar\theta)$)
along the $\bar\theta$-direction
and the anti-BRST $(s_{ab})$ and anti-co-BRST ($s_{ad}$) transformations
are equivalent to the translations
(i.e. $\mbox{Lim}_{\bar\theta \rightarrow 0} (\partial/\partial \theta)$)
along the $\theta$-direction of the supermanifold, it is straightforward
to re-express the expansion in (4.2) as follows
$$
\begin{array}{lcl} 
\tilde J_\mu (x, \theta, \bar\theta) = J_\mu (x) + \theta \; 
(s_{ab} J_\mu (x)) + \bar \theta\; (s_b J_\mu (x)) 
+ \theta\; \bar\theta\; (s_b s_{ab} J_\mu (x)).
\end{array} \eqno(4.4)
$$
It can be checked explicitly that, under the (anti-)BRST transformations (2.3),
the conserved current $J_\mu (x)$ remains invariant
(i.e. $s_{b} J_\mu (x) = s_{ab} J_\mu (x) = 0$).
This statement, with the help of (4.2) and (4.3),
 can be mathematically expressed as
$$
\begin{array}{lcl} 
b_1 \gamma_\mu \psi
= \bar \psi \gamma_\mu b_2, \qquad
\bar b_2 \gamma_\mu \psi
= \bar \psi \gamma_\mu \bar b_1, \qquad
\bar f \gamma_\mu \psi
+ \bar \psi \gamma_\mu f = i (b_1 \gamma_\mu \bar b_1 - \bar b_2
\gamma_\mu b_2). 
\end{array} \eqno(4.5)
$$
One of the possible solutions of the above restrictions, in terms
of the components  of the basic expansions in (4.1) and the
basic fields of the Lagrangian density (2.2), is
$$
\begin{array}{lcl}
&& b_1 = - e \bar \psi C, \qquad b_2 = - e C \psi,
\qquad \bar b_1 = - e \bar C \psi, \qquad \bar b_2 = - e \bar \psi \bar C,
\nonumber\\
&& f = - i e\; [\; B + e \bar C C\; ]\; \psi,
\qquad \bar f = + i e\; \bar \psi\; [\; B + e C \bar C \;].
\end{array} \eqno(4.6)
$$
At the moment, it appears to us that the above solutions are the
{\it unique} solutions to all the restrictions in (4.5)
\footnote{ Let us focus on $b_1 \gamma_\mu \psi = \bar\psi
\gamma_\mu b_2$. It is evident that the pair
of bosonic components $b_1$ and $b_2$ should be proportional
to the pair of fermionic fields $\bar\psi$ and $\psi$, respectively.
To make the latter pair bosonic in nature, we have to include the ghost
field $C$ of the Lagrangian
density (2.2) to obtain: $b_1 \sim \bar\psi C, b_2 \sim C \psi$.
Rest of the choices in (4.6) follow exactly similar kind of arguments.}. 
Ultimately, the restriction that emerges on the $(2 + 2)$-dimensional
supermanifold is
$$
\begin{array}{lcl}
\tilde J_\mu (x, \theta, \bar \theta ) = J_\mu (x).
\end{array}\eqno (4.7)
$$
Physically, the above mathematical equation implies that there is
no superspace contribution to the ordinary conserved current $J_\mu (x)$. In
other words, the transformations on the Dirac fields $\psi$ and
$\bar\psi$ (cf. (2.3)) are such that the supercurrent 
$\tilde J_\mu (x,\theta,\bar\theta)$ becomes a local composite field
$J_\mu (x) = (\bar\psi \gamma_\mu \psi) (x)$ 
{\it vis-{\'a}-vis} equation (4.4) and there is no Grassmannian
contribution to it. In a more sophisticated language, the
conservation law $\partial \cdot J = 0$ remains intact despite our 
discussions connected with the superspace and supersymmetry. It is
straightforward to check that the substitution of (4.6) into (4.1) 
leads to the following
$$
\begin{array}{lcl}
\Psi\; (x, \theta, \bar \theta) &=& \psi (x) \;+ \; \theta\; 
(s_{ab}  \psi (x))
\;+ \;\bar \theta\; (s_{b} \psi (x)) 
\;+ \;\theta \;\bar \theta \;(s_{b}\;  s_{ab} \psi (x)), 
 \nonumber\\
\bar \Psi\; (x, \theta, \bar \theta) &=& \bar \psi (x) 
\;+ \;\theta\;(s_{ab} \bar \psi (x)) \;+\bar \theta\; (s_{b} \bar \psi (x))
\;+\;\theta\;\bar \theta \;(s_{b} \; s_{ab} \bar \psi (x)).
\end{array} \eqno(4.8)
$$
This establishes the fact that the nilpotent (anti-)BRST charges $Q_{(a)b}$
are the translations generators 
$(\mbox{ Lim}_{\bar \theta \rightarrow 0}(\partial/\partial \theta))
\mbox{ Lim}_{ \theta \rightarrow 0}(\partial/\partial \bar \theta)$ 
along the $(\theta)\bar\theta$ directions
of the supermanifold. The property of the nilpotency 
(i.e. $Q_{(a)b}^2 = 0$) is encoded in the
two successive translations along the Grassmannian directions of the
supermanifold (i.e. $(\partial/\partial\theta)^2 = 
(\partial/\partial\bar\theta)^2 = 0$).

Now we shall concentrate on the derivation of the symmetry transformations 
(2.4) on the matter fields  in the framework of superfield formulation.
To this end in mind, we construct the super axial-vector current
$\tilde J^{(5)}_\mu (x,\theta,\bar\theta)$ and substitute (4.1) to obtain
$$
\begin{array}{lcl} 
\tilde J^{(5)}_\mu (x, \theta, \bar\theta) &=& \bar \Psi (x,\theta,\bar\theta)
\;\gamma_\mu \gamma_5 \;\Psi (x, \theta, \bar\theta) \nonumber\\
&=& J^{(5)}_\mu (x) + \theta \; \bar K^{(5)}_\mu (x)
+ \bar \theta\; K^{(5)}_\mu (x) + i \; \theta\; \bar\theta\; L^{(5)}_\mu (x), 
\end{array} \eqno(4.9)
$$
where the above components on the r.h.s.
can be expressed, in terms of the basic
components of the expansion in (4.1), as
$$
\begin{array}{lcl} 
&& \bar K^{(5)}_\mu (x) = i\; \bigl ( \;\bar b_2 \gamma_\mu \gamma_5 \psi -
\bar \psi \gamma_\mu \gamma_5 \bar b_1 \;\bigr ),
\qquad  K^{(5)}_\mu (x) = i \;\bigl (\; b_1 \gamma_\mu \gamma_5 \psi -
\bar \psi \gamma_\mu  \gamma_5 b_2 \;\bigr ), \nonumber\\
&& L^{(5)}_\mu (x) = \bar f \gamma_\mu \gamma_5 \psi + \bar \psi \gamma_\mu 
\gamma_5 f
+ i (\bar b_2 \gamma_\mu \gamma_5 b_2 - b_1 \gamma_\mu \gamma_5 \bar b_1), 
\qquad J^{(5)}_\mu (x) = \bar \psi  \gamma_\mu \gamma_5 \psi.
 \end{array} \eqno(4.10)
$$
Invoking the analogue of the condition (4.7) (i.e. $\tilde J^{(5)}_\mu
(x,\theta, \bar\theta) = J^{(5)}_\mu (x)$), we obtain the following
conditions on the components of the super expansion in (4.9):
$$
\begin{array}{lcl}
K^{(5)}_\mu (x) = 0, \qquad 
\bar K^{(5)}_\mu (x) = 0, \qquad 
L^{(5)}_\mu (x) = 0. 
\end{array} \eqno(4.11)
$$
Ultimately,  these conditions lead to 
$$
\begin{array}{lcl}
&& b_1 = + e \bar \psi \bar C \gamma_5, \qquad 
b_2 = - e \bar C \gamma_5 \psi,
\qquad \bar b_1 = - e C \gamma_5 \psi, 
\qquad \bar b_2 = + e \bar \psi C \gamma_5,
\nonumber\\
&& f = + i e\; [ \;{\cal B} \gamma_5 - e C \bar C \;] \;\psi,
\qquad \bar f = + i e\; \bar \psi\; [\;{\cal B}\gamma_5 +
 e \bar C C \;].
\end{array} \eqno(4.12)
$$
The substitution of the above values in the super expansion in (4.1)
leads to the analogous expansion as in (4.8) with the replacements:
$s_b \rightarrow s_d, \; s_{ab}\rightarrow s_{ad}$. Thus, we obtain
$$
\begin{array}{lcl}
\Psi\; (x, \theta, \bar \theta) &=& \psi (x) \;+ \; \theta\; 
(s_{ad}  \psi (x))
\;+ \;\bar \theta\; (s_{d} \psi (x)) 
\;+ \;\theta \;\bar \theta \;(s_{d}\;  s_{ad} \psi (x)), 
 \nonumber\\
\bar \Psi\; (x, \theta, \bar \theta) &=& \bar \psi (x) 
\;+ \;\theta\;(s_{ad} \bar \psi (x)) \;+\bar \theta\; (s_{d} \bar \psi (x))
\;+\;\theta\;\bar \theta \;(s_{d} \; s_{ad} \bar \psi (x)).
\end{array} \eqno(4.13)
$$
This provides the
 geometrical interpretation for the (anti-)co-BRST charges as
the translation generators along the $(\theta)\bar\theta$ directions
of the supermanifold. This interpretation is exactly identical to
the interpretation for the (anti-)BRST charges as the translation generators.
The above statement for the (anti-)BRST- and (anti-)co-BRST charges
can be succinctly expressed in the mathematical form, using (2.5), as 
$$
\begin{array}{lcl}
s_{r} \Sigma (x) = \mbox{Lim}_{\theta \rightarrow 0} 
{\displaystyle \frac{\partial}{\partial
\bar\theta}} \tilde \Sigma (x,\theta,\bar\theta)
\equiv - i \{ \Sigma (x), Q_{r} \}, \nonumber\\
s_{t} \Sigma (x) = \mbox{Lim}_{\bar \theta \rightarrow 0} 
{\displaystyle \frac{\partial}{\partial
\theta}} \tilde \Sigma (x,\theta,\bar\theta)
\equiv - i \{ \Sigma (x), Q_{t} \}, 
\end{array} \eqno(4.14)
$$
where $r = b, d$,  $\;t = ab, ad$ and $\Sigma (x) = \psi (x), \bar \psi (x),\;
\tilde \Sigma (x,\theta,\bar\theta) = \Psi (x,\theta, \bar\theta), 
\bar \Psi (x,\theta,\bar\theta)$. Thus, it is clear
that the mapping that exists
among the symmetry transformations, the conserved charges and 
the translation generators along the Grassmannian directions are
$$
\begin{array}{lcl}
s_{b(d)} \leftrightarrow Q_{b(d)} \leftrightarrow
\mbox{Lim}_{\theta \rightarrow 0} {\displaystyle \frac{\partial}{\partial
\bar\theta}},\;\; 
s_{ad} \leftrightarrow Q_{ad} \leftrightarrow
\mbox{Lim}_{\bar \theta \rightarrow 0} {\displaystyle \frac{\partial}{\partial
\theta}},\;\; 
s_{ab} \leftrightarrow Q_{ab} \leftrightarrow
\mbox{Lim}_{\bar\theta \rightarrow 0} {\displaystyle \frac{\partial}{\partial
\theta}}.
\end{array}\eqno(4.15)
$$

\noindent
{\bf 5 Conclusions}\\

\noindent
In the present investigation, we set out to derive the 
off-shell nilpotent (anti-)BRST
and (anti-)co-BRST symmetries for the matter (Dirac) fields in the framework
of geometrical superfield approach to BRST formalism
We chose the two-dimensional
interacting $U(1)$ gauge theory (i.e. QED) for our discussion primarily
for two reasons. First and foremost, this theory provides one of
the simplest gauge theory and 
a {\it unique interacting}  field theoretical model for the Hodge theory.
Second, the Lagrangian density (2.2) of this theory is endowed with
a local, covariant, continuous and nilpotent (anti-)co-BRST symmetries
which is not the case for the four dimensional QED where the (anti-)co-BRST
transformations are non-local and non-covariant (see, e.g., [31] for
details). We have been able to derive the off-shell nilpotent
(anti-)BRST and (anti-)co-BRST
symmetry transformations on the Dirac fields by invoking a couple of
restrictions (i.e. $ \tilde J_\mu (x,\theta,\bar\theta) = J_\mu (x)$ and 
$\tilde J_\mu^{(5)}
(x,\theta,\bar\theta) = J_\mu^{(5)} (x)$) on the $(2 + 2)$-dimensional
supermanifold. In contrast to the 
(dual-)horizontality conditions, these restrictions are not imposed 
by hand from the outside. Rather,
they appear very naturally because of the fact that
$s_{(a)b} J_\mu (x) = 0, s_{(a)d} J_\mu^{(5)} (x) = 0$ in the super expansion 
of the super currents $\tilde J_\mu (x,\theta,\bar\theta)$ and
$\tilde J_\mu^{(5)} (x,\theta,\bar\theta)$ (cf. eqns. (4.4) and (4.9)).
Physically, these conditions imply nothing but the conservation of the
electric charge for the massive Dirac fields and the conservation of
the spin (i.e. helicity in 2D spacetime) 
for the massless Dirac fields, respectively. These conservation laws 
persist even in the superfield formulation of the theory.
This is why, automatically, we get the conditions
$\tilde J_\mu (x,\theta,\bar\theta) = J_\mu (x)$ and
$\tilde J^{(5)}_\mu (x,\theta,\bar\theta) = J^{(5)}_\mu (x)$. We would
like to comment that our method of derivation of the (anti-)BRST 
transformations for the matter fields, in the framework of the superfield
formalism, can be generalized to the physical 4D Abelian as well as non-Abelian 
gauge theories (see, e.g., [31,32] for transformations).
It would be also interesting to obtain the on-shell
nilpotent version of the above symmetries in the framework of the superfield
formulation. These are some of the open problems which are under investigation
and our results would be reported elsewhere [33].

\end{document}